\pdfoutput=1
\RequirePackage{fix-cm}
\documentclass[reqno]{amsproc}
\usepackage{amssymb}
\usepackage{hyperref}
\usepackage{microtype}
\usepackage{lmodern}

\usepackage{tikz}
\usetikzlibrary{decorations.pathreplacing}

\usepackage{graphicx}

\usepackage[citation-order,nobysame]{amsrefs}
\usepackage{xyzbib}

\usepackage{mathtools}
\mathtoolsset{showonlyrefs}
\allowdisplaybreaks

\numberwithin{equation}{section}
\let\cite=\cites

\newtheorem{theorem}{Theorem}
\newtheorem{conjecture}{Conjecture}
\DeclareMathOperator*{\res}{\mathrm{res}}
\DeclareMathOperator*{\diag}{\mathrm{diag}}
\DeclareMathOperator*{\ad}{\mathrm{ad}}

\let\geq=\geqslant

\newcommand{\bra}[1]{\langle #1 \rvert}
\newcommand{\ket}[1]{\lvert #1 \rangle}
\newcommand\bracket[2]{\langle #1\vert #2\rangle}

\begin{document}


\title{Periodic Motzkin chain: Ground states and symmetries}

\author{Andrei G. Pronko}
\address{Steklov Mathematical Institute, 
Fontanka 27, 191023 Saint Petersburg, Russia}
\email{a.g.pronko@gmail.com}

\begin{abstract}  
Motzkin chain is a model of nearest-neighbor interacting 
quantum $s=1$ spins with open boundary conditions. It is known  
that it has a unique ground state which can be viewed 
as a sum of Motzkin paths. 
We consider the case of periodic boundary conditions and 
provide several conjectures about 
structure of the ground state space and symmetries of the Hamiltonian. 
We conjecture that the ground state is degenerate 
and independent states are distinguished by 
eigenvalues of the third component of total spin operator. 
Each of these states can be described as a sum of paths, 
similar to the Motzkin paths.  
Moreover, there exist two operators commuting with the Hamiltonian,  
which play the roles of lowering and raising operators when acting 
at these states. 
We conjecture also that these operators generate a $C$-type Lie 
algebra, with rank equal to the number of sites. 
The symmetry algebra of the Hamiltonian is actually wider, and
extended, besides the cyclic shift operator, 
by a central element contained in the  
third component of total spin operator.  

\end{abstract}

\maketitle
\section{Introduction}

Motzkin spin chain originally appeared in 
the context of study of long-range entanglement in the ground states
of critical one-dimensional quantum systems \cite{BCMNS-12}. 
It is an example of open spin-$1$ chain exhibiting criticality without 
a  ``frustration'': the model is gapless and its  
unique ground state minimizes all individual terms of the Hamiltonian.
The later property ensures stability of the ground state, 
which can be viewed as the uniform superposition 
of Motzkin paths, against 
possible inclusion of term-dependent interaction parameters 
in the Hamiltonian.   
This ground state, called Motzkin state, has one more interesting 
feature: entanglement entropy grows logarithmically as the   
size of a subsystem increases, instead of being bounded 
by a constant that occurs in critical spin-$1/2$ chains. 
Similar properties have been found for ground states 
of higher integer spin generalizations, or ``colored'' versions, of 
the Motzkin  chain \cite{MS-16}, and half-integer spin chains with 
interaction of three nearest neighbors \cite{SK-17}. For 
advances in study of these models and further references, 
see \cite{AAZK-19,CBG-24,MGM-24}. 

A sensible problem which can be addressed is whether the Motzkin
chain is an integrable quantum model. Although 
its ground state is known exactly, exact information  
is still lacking  about its excited states, 
that is important, in particular in the study of 
correlation functions. More specifically, one could be interested in 
finding structures closely related to integrability of the model, such as 
underlying Yang--Baxter relation and a quantum transfer matrix 
generating the Hamiltonian \cite{TTF-83,S-82,KBI-93,S-20}. This indeed turns
out to be possible for a ``free'' version of the Motzkin chain, in which 
one of the terms describing interaction of spins in the Hamiltonian density 
is omitted \cite{HSK-23}.
One can also notice a raised recent interest in search of new solutions 
for the Yang--Baxter equation, see, e.g., \cite{K-23,PK-24,SSPK-24,GBFG-24}, 
motivated by possible applications in the theory of quantum computations.

In this paper, the Motzkin chain is considered 
for periodic boundary conditions.  
Our results are purely conjectural and obtained by studying systems of small 
size, up to six sites in the length, we present details for up to 
the four-site case. We find that 
the ground state of the chain with $N$ sites 
is $2N+1$ times degenerated, 
with independent states distinguished by the eigenvalue of 
the third component of total spin 
operator, $S^z=0,\pm 1,\dots, \pm N$. Moreover, these states also admit an
interpretation in terms of paths, similar to the Motzkin paths 
but less restricted. The degeneracy of the ground state   
hints at existence of a quantum symmetry algebra, namely, a set of operators 
commuting with the Hamiltonian. We conjecture explicit formulas 
for operators which admit interpretations as lowering and raising   
operators when acting at the ground states. Furthermore, 
we conjecture that they generate the Lie algebra $C_N$. 
The symmetry algebra of the Hamiltonian is actually wider, 
extended by the cyclic shift operator and 
a central element contained in the 
third component of total spin operator along with  
elements of the Cartan subalgebra of $C_N$.

Although our results do not immediately 
imply quantum integrability of the periodic 
Motzkin chain, they may prove useful for searching 
suitable algebraic structures among 
solutions of the Yang--Baxter equation. Such structures, once 
identified, could provide proofs 
of our conjectures and be applied for construction of other models 
with similar properties, 
that may represent an independent interest.

The paper is organized as follows. In the next section 
we recall origin of the open Motzkin chain. 
In section 3 we consider the periodic
Motzkin chain and formulate four conjectures about 
the ground states and symmetries.
In section 4 we give some details for 
two-, three-, and four-site chains. 

\section{Motzkin paths, Motzkin state, and open chain}

In this section we mainly introduce the notation and recall an origin
of the (open) Motzkin chain. 
 
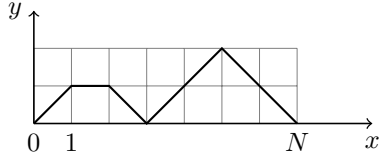
\begin{figure}
\centering

\begin{tikzpicture}[scale=.5]

\draw[help lines] (0,0) grid (7,2);


\draw [thick] (0,0)--(1,1)--(2,1)--(3,0)--(4,1)--(5,2)--(6,1)--(7,0);

\draw [semithick] [->] (0,0)--(9,0);
\draw [semithick] [->] (0,0)--(0,3);

\node at (0,-.5) {$0$};
\node at (9,-.5) {$x$};
\node at (-.5,3) {$y$};
\node at (1,-.5) {$1$};
\node at (7,-.5) {$N$};

\end{tikzpicture}	

\caption{A Motzkin path, $N=7$}
\label{fig-MotzkinPath}
\end{figure}

Consider square lattice with vertices labeled by $(x,y)$ 
with $x$ and $y$ being integers. A Motzkin path of $N$ steps starts
at $(x,y)=(0,0)$ and ends at $(x,y)=(0,N)$, 
steps are made along the $x$-direction, at each step $\Delta x=1$ and 
$\Delta y \in \{-1,0,1\}$, with the restriction that $y\geq 0$ 
along the path. 
An example of Motzkin path is shown in 
\figurename~\ref{fig-MotzkinPath}.

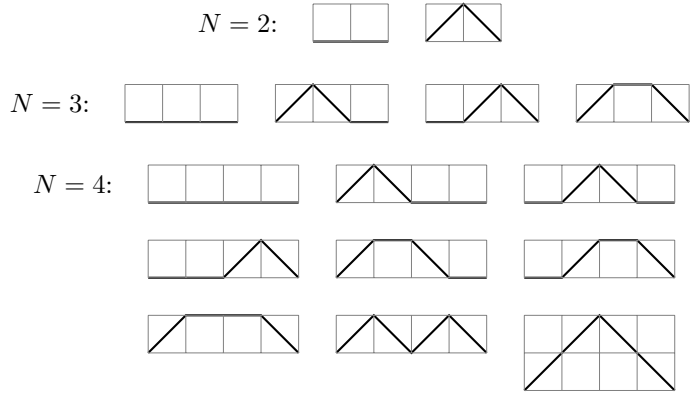
\begin{figure}
\centering

\begin{tikzpicture}[scale=.5]
\node at (-2,.5) {$N=2$:};
\draw [thick] (0,0)--(2,0);
\draw [help lines] (0,0) grid (2,1);
\draw [thick] (3,0)--(4,1)--(5,0);
\draw [help lines] (3,0) grid (5,1);
\end{tikzpicture}	
\vspace{.2in}\par
\begin{tikzpicture}[scale=.5]
\node at (-2,.5) {$N=3$:};
\draw [thick] (0,0)--(3,0);
\draw [help lines] (0,0) grid (3,1);
\draw [thick] (4,0)--(5,1)--(6,0)--(7,0);
\draw [help lines] (4,0) grid (7,1);
\draw [thick] (8,0)--(9,0)--(10,1)--(11,0);
\draw [help lines] (8,0) grid (11,1);
\draw [thick] (12,0)--(13,1)--(14,1)--(15,0);
\draw [help lines] (12,0) grid (15,1);
\end{tikzpicture}	
\vspace{.2in}\par
\begin{tikzpicture}[scale=.5]
\node at (-2,.5) {$N=4$:};
\draw [thick] (0,0)--(4,0);
\draw [help lines] (0,0) grid (4,1);
\draw [thick] (5,0)--(6,1)--(7,0)--(8,0)--(9,0);
\draw [help lines] (5,0) grid (9,1);
\draw [thick] (10,0)--(11,0)--(12,1)--(13,0)--(14,0);
\draw [help lines] (10,0) grid (14,1);
\draw [thick] (0,-2)--(2,-2)--(3,-1)--(4,-2);
\draw [help lines] (0,-2) grid (4,-1);
\draw [thick] (5,-2)--(6,-1)--(7,-1)--(8,-2)--(9,-2);
\draw [help lines] (5,-2) grid (9,-1);
\draw [thick] (10,-2)--(11,-2)--(12,-1)--(13,-1)--(14,-2);
\draw [help lines] (10,-2) grid (14,-1);
\draw [thick] (0,-4)--(1,-3)--(2,-3)--(3,-3)--(4,-4);
\draw [help lines] (0,-4) grid (4,-3);
\draw [thick] (5,-4)--(6,-3)--(7,-4)--(8,-3)--(9,-4);
\draw [help lines] (5,-4) grid (9,-3);
\draw [thick] (10,-5)--(12,-3)--(14,-5);
\draw [help lines] (10,-5) grid (14,-3);
\end{tikzpicture}	

\caption{Motzkin paths for $N=2,3,4$}
\label{fig-OpenChain}
\end{figure}

We denote by $\mathcal{M}_N$ the set of Motzkin paths of length $N$. 
The number of elements in $\mathcal{M}_N$ is known as a Motzkin number, 
and its original definition is the number of all possible 
non-intersecting chords of a circle with $N$ nodes. For 
$N=1,2,3,4,5,\dots$ the Motzkin numbers form 
the sequence $1,2,4,9,21,\dots$, see \cite{A001006}, where a detailed 
description of various interpretations and related references 
can also be found. Examples of paths for $N=2,3,4$ are shown 
in \figurename~\ref{fig-OpenChain}. 

To connect Motzkin paths with a quantum spin chain, let us denote  
basis vectors in $\mathbb{C}^3$ by
\begin{equation}
\ket{\mathrm{u}}=
\begin{pmatrix}
1 \\ 0 \\ 0
\end{pmatrix},
\qquad 
\ket{\mathrm{f}}=
\begin{pmatrix}
0 \\ 1 \\ 0
\end{pmatrix},
\qquad 
\ket{\mathrm{d}}=
\begin{pmatrix}
0 \\ 0 \\ 1
\end{pmatrix}.
\end{equation}
In what follows we associate these vectors with up, flat (or forward), 
and down steps in lattice paths. For further use we also introduce 
spin-$1$ matrices
\begin{equation}\label{spin1rep}
s^{+}=
\begin{pmatrix}
0 & 1 & 0 \\ 0 & 0 & 1 \\ 0 & 0 & 0  
\end{pmatrix},\qquad
s^{-}=
\begin{pmatrix}
0 & 0 & 0 \\ 1 & 0 & 0 \\ 0 & 1 & 0  
\end{pmatrix},\qquad
s^{z}=
\begin{pmatrix}
1 & 0 & 0 \\ 0 & 0 & 0 \\ 0 & 0 & -1  
\end{pmatrix}.
\end{equation}
These matrices correspond to the vector representation 
of $\mathfrak{sl}_2$ with the exception that usually 
$s^\pm$ are defined with the factor $1/\sqrt{2}$; 
for our purposes below it will be convenient that $s^\pm$ are 
build of $0$ and $1$.   

Let us now consider the vector space $(\mathbb{C}^3)^{\otimes N}$. 
For the basis vectors we use the notation
\begin{equation}
\ket{\ell_1\ell_2\dots\ell_N}=\ket{\ell_1}\otimes \ket{\ell_2}
\otimes\dots\otimes \ket{\ell_N}, \qquad \ell_1,\ell_2,\dots,\ell_N=
\mathrm{u,f,d}.
\end{equation}
The space $(\mathbb{C}^3)^{\otimes N}$ 
can be regarded as a Hilbert space of a quantum spin-1 chain 
with $N$ sites, with $j$th factor $\mathbb{C}^3$ in the direct product 
treated as $j$th site of the chain. 
We use the standard notation $s_j^{\pm,z}$
for the local 
spin operators acting in $j$th copy of $\mathbb{C}^3$, that is
\begin{equation}\label{spmsz}
s_j^{\pm,z}=I^{\otimes j-1}\otimes s^{\pm,z}\otimes I^{\otimes N-j},
\end{equation}     
where $I$ stands for $3\times 3$ identity matrix. An important object 
appearing below is the ``third'' component of 
total spin operator: 
\begin{equation}\label{Szop}
\mathrm{S}^z= \sum_{j=1}^N s_j^z.
\end{equation}
This is a diagonal matrix with the eigenvalues $S^z=0,\pm 1,\dots,\pm N$ 
(we denote by upright capital letters ``global'' operators and by  
their italic variants the eigenvalues of these operators).

Given a Motzkin path, a vector in $(\mathbb{C}^3)^{\otimes N}$ can be 
associated to this path according to the steps made to produce it, 
with up, flat, and 
down steps corresponding to the letters 
$\ell=\mathrm{u,f,d}$ in the 
vector notation. For example, the path 
shown in \figurename~\ref{fig-MotzkinPath} consists of steps 
``$\mathrm{ufduudd}$'', hence
the corresponding vector is $\ket{\mathrm{ufduudd}}$. 

Having the set $\mathcal{M}_N$ of Motzkin paths of length $N$, 
so-called Motzkin state $\ket{\mathcal{M}_N}$ can be defined 
as the uniform sum over elements of this set:
\begin{equation}
\ket{\mathcal{M}_N}=
\sum_{\ell_1 \ell_2\dots \ell_N\in \mathcal{M}_N}
\ket{\ell_1\ell_2\dots\ell_N}. 
\end{equation}
The Motzkin states for $N=2,3,4$, 
see \figurename~\ref{fig-OpenChain}, are 
\begin{align}
\ket{\mathcal{M}_2}&=\ket{\mathrm{ff}}+\ket{\mathrm{ud}},
\\
\ket{\mathcal{M}_3}&=\ket{\mathrm{fff}}+\ket{\mathrm{udf}}
+\ket{\mathrm{fud}}+\ket{\mathrm{ufd}},
\\
\ket{\mathcal{M}_4}&=\ket{\mathrm{ffff}}+\ket{\mathrm{udff}}
+\ket{\mathrm{fudf}}
+\ket{\mathrm{ffud}}
\\ &\quad
+\ket{\mathrm{ufdf}}+\ket{\mathrm{fufd}}
+\ket{\mathrm{uffd}}+\ket{\mathrm{udud}}+\ket{\mathrm{uudd}}.
\end{align}
The Motzkin state is an eigenstate of the operator 
$\mathrm{S}^z$ with zero eigenvalue,
\begin{equation}
\mathrm{S}^z\ket{\mathcal{M}_N}=0. 
\end{equation}

An interesting and important feature of the Motzkin state is that there exists 
a Hamiltonian with nearest-neighbor interaction having 
it as a unique ground state \cite{BCMNS-12}. 
The nearest-neighbor interaction is described by the three-dimensional 
projector 
\begin{equation}
\Pi=\mathrm{U}+\mathrm{D}+\mathrm{F},\qquad 
\mathrm{U},\mathrm{D},
\mathrm{F}\in \text{End}(\mathbb{C}^3\otimes \mathbb{C}^3),
\end{equation}
where $\mathrm{U}$, $\mathrm{D}$, and $\mathrm{F}$ are mutually 
commuting and orthogonal to each other 
one-dimensional projectors
\begin{equation}\label{UDF}
\begin{split}
\mathrm{U}&=\frac{1}{2}
\left(\ket{\mathrm{uf}}-\ket{\mathrm{fu}}\right)
\left(\bra{\mathrm{uf}}-\bra{\mathrm{fu}}\right),
\\
\mathrm{D}&=\frac{1}{2}
\left(\ket{\mathrm{df}}-\ket{\mathrm{fd}}\right)
\left(\bra{\mathrm{df}}-\bra{\mathrm{fd}}\right),
\\
\mathrm{F}&=\frac{1}{2}
\left(\ket{\mathrm{ud}}-\ket{\mathrm{ff}}\right)
\left(\bra{\mathrm{ud}}-\bra{\mathrm{ff}}\right).
\end{split}
\end{equation}
In $9\times 9$ matrix realization of operators acting in 
$\mathbb{C}^3\otimes \mathbb{C}^3$ (as $3\times 3$ 
block matrices with respect to the first factor, with entries 
in the $3\times 3$ blocks corresponding to the second factor)  
this projector reads
\begin{equation}\label{Pi}
\Pi=\frac{1}{2}
\begin{pmatrix}
 0 & 0 & 0 & 0 & 0 & 0 & 0 & 0 & 0 \\
 0 & 1 & 0 & -1 & 0 & 0 & 0 & 0 & 0 \\
 0 & 0 & 1 & 0 & -1 & 0 & 0 & 0 & 0 \\
 0 & -1 & 0 & 1 & 0 & 0 & 0 & 0 & 0 \\
 0 & 0 & -1 & 0 & 1 & 0 & 0 & 0 & 0 \\
 0 & 0 & 0 & 0 & 0 & 1 & 0 & -1 & 0 \\
 0 & 0 & 0 & 0 & 0 & 0 & 0 & 0 & 0 \\
 0 & 0 & 0 & 0 & 0 & -1 & 0 & 1 & 0 \\
 0 & 0 & 0 & 0 & 0 & 0 & 0 & 0 & 0 \\
\end{pmatrix}.
\end{equation}
The following result is our starting point. 
\begin{theorem}[Bravyi, Caha, Movassagh, Nagaj, Shor \cite{BCMNS-12}]
The Motzkin state 
$\ket{\mathcal{M}_N}$ is a unique ground state with zero eigenvalue 
of the spin chain Hamiltonian
\begin{equation}\label{Hopen}
\mathrm{H}=
\sum_{i=1}^{N-1} \Pi_{i,i+1}+
\ket{\mathrm{d}}\bra{\mathrm{d}}_1+\ket{\mathrm{u}}\bra{\mathrm{u}}_N,
\end{equation}
where subscripts indicate sites where operators act. 
\end{theorem}

Hamiltonian \eqref{Hopen} commutes with the operator \eqref{Szop},
and no other operators commuting with the Hamiltonian are known.
For this reason, the Motzkin chain in its original formulation with 
open boundary conditions is believed to be a 
non-integrable quantum model, although its ground state
is known exactly (given by the Motzkin state).

\section{Periodic Motzkin chain}

Given open spin chain Hamiltonian \eqref{Hopen}, one may wonder about 
properties of its periodic version, where the boundary term is chosen 
such that the Hamiltonian is cyclic invariant, 
\begin{equation}\label{Hpbc}
\mathrm{H}^\text{Periodic}=
\sum_{i=1}^{N-1} \Pi_{i,i+1}+\Pi_{N,1}.
\end{equation}
 
A simple albeit not typical feature of the Hamiltonian density 
of the Motzkin chain consists in $\Pi$ acting non-symmetrically at factors 
in $\mathbb{C}^3\otimes \mathbb{C}^3$; indeed, 
the symmetry is broken by the $\mathrm{F}$ term in \eqref{UDF}.
This means that $[\Pi,\mathrm{P}]\ne 0$, where $\mathrm{P}$ is 
the permutation operator in $\mathbb{C}^3\otimes \mathbb{C}^3$, 
\begin{equation}\label{Pop}
\mathrm{P}=
\begin{pmatrix}
 1 & 0 & 0 & 0 & 0 & 0 & 0 & 0 & 0 \\
 0 & 0 & 0 & 1 & 0 & 0 & 0 & 0 & 0 \\
 0 & 0 & 0 & 0 & 0 & 0 & 1 & 0 & 0 \\
 0 & 1 & 0 & 0 & 0 & 0 & 0 & 0 & 0 \\
 0 & 0 & 0 & 0 & 1 & 0 & 0 & 0 & 0 \\
 0 & 0 & 0 & 0 & 0 & 0 & 0 & 1 & 0 \\
 0 & 0 & 1 & 0 & 0 & 0 & 0 & 0 & 0 \\
 0 & 0 & 0 & 0 & 0 & 1 & 0 & 0 & 0 \\
 0 & 0 & 0 & 0 & 0 & 0 & 0 & 0 & 1 \\
\end{pmatrix}.
\end{equation}
We note that the last term in \eqref{Hpbc} can be written as 
a product of operators acting only at nearest-neighbor sites: 
\begin{equation}\label{PiN1}
\Pi_{N,1}=\mathrm{P}_{N-1,N}\mathrm{P}_{N-2,N-1}\cdots \mathrm{P}_{1,2} 
\Pi_{1,2} \mathrm{P}_{1,2} \cdots \mathrm{P}_{N-2,N-1} \mathrm{P}_{N-1,N}. 
\end{equation} 
This formula, together with \eqref{Pi} and \eqref{Pop},
is useful when the Hamiltonian \eqref{Hpbc} 
need to be considered as a matrix in symbolic computer calculations. 

The strings of permutation operators standing 
at both sides of $\Pi_{1,2}$ in \eqref{PiN1} are nothing but 
the cyclic shift operator $\mathrm{C}$ acting at sites as 
$i\mapsto i+1$ and its inverse $\mathrm{C}^{-1}$ acting at sites 
as $i\mapsto i-1$, 
\begin{equation}\label{Cop}
\mathrm{C}=\mathrm{P}_{1,2} \cdots \mathrm{P}_{N-2,N-1} \mathrm{P}_{N-1,N}.
\end{equation}
This is an integral of motion, 
\begin{equation}
\left[\mathrm{C},\mathrm{H}^\text{Periodic}\right]=0.
\end{equation}
Note that the operator $\mathrm{S}^z$ 
commutes with the Hamiltonian \eqref{Hpbc}.

Now we turn to the results. They are summarized in four conjectures. 
The first two conjectures are obtained by exact symbolic manipulations 
for the model up to $N=6$ case, the remaining two have been 
obtained for up to $N=4$ case and verified numerically in the $N=5$ case.   
In the next section we consider 
examples of $N=2,3,4$ in some detail; 
here we focus on formulating the results in 
general. 

The first conjecture is about the structure 
of the space of the ground state, which is essentially 
the null space of the Hamiltonian ($H^\text{periodic}=0$).    

\begin{conjecture}
The periodic Motzkin chain with $N$ sites has $2N+1$ degenerate 
ground state with zero eigenvalue, with independent states  
$\ket{v_{S^z}}$  
labeled by eigenvalues of the third component of total spin operator, 
$S^z=0,\pm 1,\dots, \pm N$. These 
states can be described as sums of paths connecting points
$(x,y)=(0,0)$ and $(x,y)=(N,S^z)$, having at each step $\Delta x=1$ 
and $\Delta y\in \{-1,0,1\}$, with no restriction on the value of $y$ along the path.
\end{conjecture}

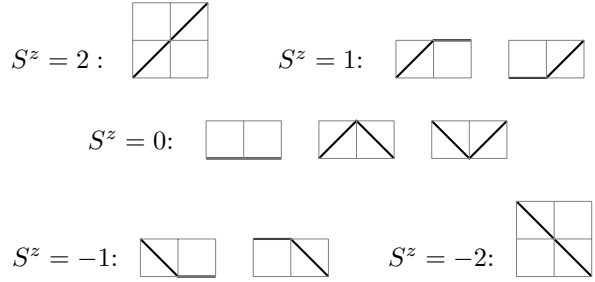
\begin{figure}
\centering

\begin{tikzpicture}[scale=.5]
\node at (-2,.5) {$S^z=2:$};
\draw [thick] (0,0)--(2,2);
\draw [help lines] (0,0) grid (2,2);
\node at (5,.5) {$S^z=1$:};
\draw [thick] (7,0)--(8,1)--(9,1); 
\draw [help lines] (7,0) grid (9,1);
\draw [thick] (10,0)--(11,0)--(12,1); 
\draw [help lines] (10,0) grid (12,1);
\end{tikzpicture}	
\vspace{.2in}\par
\begin{tikzpicture}[scale=.5]
\node at (-2,.5) {$S^z=0$:};
\draw [thick] (0,0)--(2,0);
\draw [help lines] (0,0) grid (2,1);
\draw [thick] (3,0)--(4,1)--(5,0);
\draw [help lines] (3,0) grid (5,1);
\draw [thick] (6,1)--(7,0)--(8,1);
\draw [help lines] (6,0) grid (8,1);
\end{tikzpicture}	
\vspace{.2in}\par
\begin{tikzpicture}[scale=.5]
\node at (-2,-.5) {$S^z=-1$:};
\draw [thick] (0,0)--(1,-1)--(2,-1);
\draw [help lines] (0,-1) grid (2,0);
\draw [thick] (3,0)--(4,0)--(5,-1);
\draw [help lines] (3,-1) grid (5,0);
\node at (8,-.5) {$S^z=-2$:};
\draw [thick] (10,1)--(12,-1);
\draw [help lines] (10,-1) grid (12,1);
\end{tikzpicture}	

\caption{Paths connecting points $(x,y)=(0,0)$ and
$(x,y)=(N,S^z)$ corresponding to the five ground states of the
periodic two-site chain}
\label{fig-TwoSiteGS}
\end{figure}

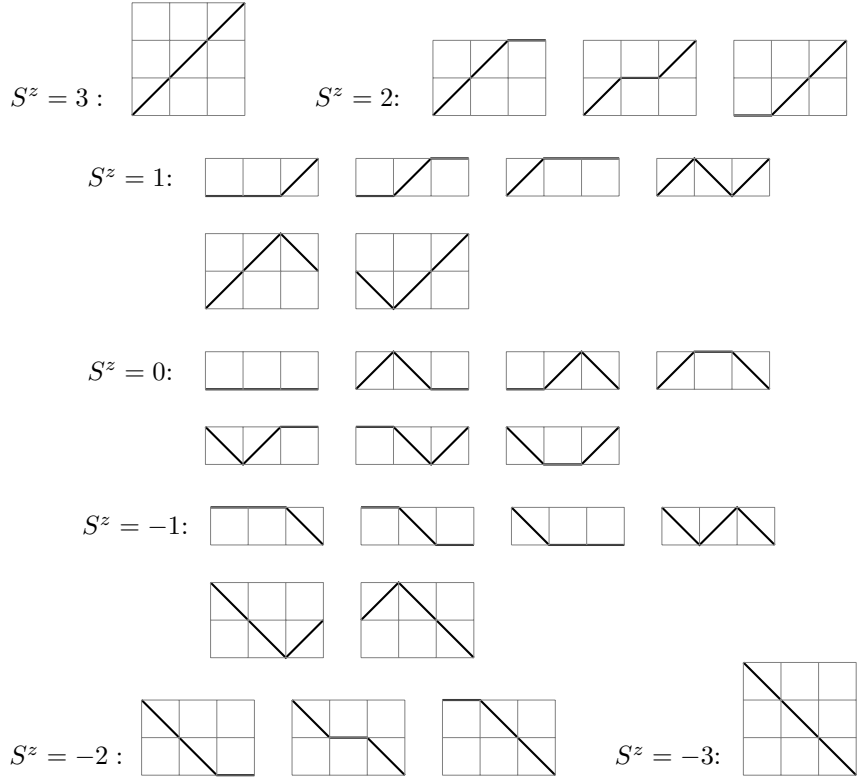
\begin{figure}
\centering

\begin{tikzpicture}[scale=.5]
\node at (-2,.5) {$S^z=3:$};
\draw [thick] (0,0)--(3,3);
\draw [help lines] (0,0) grid (3,3);
\node at (6,.5) {$S^z=2$:};
\draw [thick] (8,0)--(10,2)--(11,2); 
\draw [help lines] (8,0) grid (11,2);
\draw [thick] (12,0)--(13,1)--(14,1)--(15,2); 
\draw [help lines] (12,0) grid (15,2);
\draw [thick] (16,0)--(17,0)--(19,2); 
\draw [help lines] (16,0) grid (19,2);
\end{tikzpicture}	
\vspace{.2in}\par
\begin{tikzpicture}[scale=.5]
\node at (-2,.5) {$S^z=1$:};
\draw [thick] (0,0)--(2,0)--(3,1);
\draw [help lines] (0,0) grid (3,1);
\draw [thick] (4,0)--(5,0)--(6,1)--(7,1);
\draw [help lines] (4,0) grid (7,1);
\draw [thick] (8,0)--(9,1)--(11,1);
\draw [help lines] (8,0) grid (11,1);
\draw [thick] (12,0)--(13,1)--(14,0)--(15,1);
\draw [help lines] (12,0) grid (15,1);
\draw [thick] (0,-3)--(2,-1)--(3,-2);
\draw [help lines] (0,-3) grid (3,-1);
\draw [thick] (4,-2)--(5,-3)--(7,-1);
\draw [help lines] (4,-3) grid (7,-1);
\end{tikzpicture}	
\vspace{.2in}\par
\begin{tikzpicture}[scale=.5]
\node at (-2,.5) {$S^z=0$:};
\draw [thick] (0,0)--(3,0);
\draw [help lines] (0,0) grid (3,1);
\draw [thick] (4,0)--(5,1)--(6,0)--(7,0);
\draw [help lines] (4,0) grid (7,1);
\draw [thick] (8,0)--(9,0)--(10,1)--(11,0);
\draw [help lines] (8,0) grid (11,1);
\draw [thick] (12,0)--(13,1)--(14,1)--(15,0);
\draw [help lines] (12,0) grid (15,1);
\draw [thick] (0,-1)--(1,-2)--(2,-1)--(3,-1);
\draw [help lines] (0,-2) grid (3,-1);
\draw [thick] (4,-1)--(5,-1)--(6,-2)--(7,-1);
\draw [help lines] (4,-2) grid (7,-1);
\draw [thick] (8,-1)--(9,-2)--(10,-2)--(11,-1);
\draw [help lines] (8,-2) grid (11,-1);
\end{tikzpicture}	
\vspace{.2in}\par
\begin{tikzpicture}[scale=.5]
\node at (-2,.5) {$S^z=-1$:};
\draw [thick] (0,1)--(2,1)--(3,0);
\draw [help lines] (0,0) grid (3,1);
\draw [thick] (4,1)--(5,1)--(6,0)--(7,0);
\draw [help lines] (4,0) grid (7,1);
\draw [thick] (8,1)--(9,0)--(11,0);
\draw [help lines] (8,0) grid (11,1);
\draw [thick] (12,1)--(13,0)--(14,1)--(15,0);
\draw [help lines] (12,0) grid (15,1);
\draw [thick] (0,-1)--(2,-3)--(3,-2);
\draw [help lines] (0,-3) grid (3,-1);
\draw [thick] (4,-2)--(5,-1)--(7,-3);
\draw [help lines] (4,-3) grid (7,-1);
\end{tikzpicture}	
\vspace{0in}\par
\begin{tikzpicture}[scale=.5]
\node at (-2,.5) {$S^z=-2:$};
\draw [thick] (0,2)--(2,0)--(3,0); 
\draw [help lines] (0,0) grid (3,2);
\draw [thick] (4,2)--(5,1)--(6,1)--(7,0); 
\draw [help lines] (4,0) grid (7,2);
\draw [thick] (8,2)--(9,2)--(11,0); 
\draw [help lines] (8,0) grid (11,2);
\node at (14,.5) {$S^z=-3$:};
\draw [thick] (16,3)--(19,0);
\draw [help lines] (16,0) grid (19,3);
\end{tikzpicture}	

\caption{Paths connecting points $(x,y)=(0,0)$ and
$(x,y)=(N,S^z)$ corresponding to the seven ground states of the
periodic three-site chain}
\label{fig-ThreeSiteGS}
\end{figure}

Paths describing the ground states  $\ket{v_{S^z}}$
in the case of $N=2$ are shown in \figurename~\ref{fig-TwoSiteGS}.
The case $N=3$ is considered in 
\figurename~\ref{fig-ThreeSiteGS}.
Note that the states  $\ket{v_{S^z}}$ are cyclic invariant, i.e., 
eigenstates of 
the cyclic shift operator with unit eigenvalue:
\begin{equation}
\mathrm{C} \ket{v_{S^z}}= \ket{v_{S^z}}.
\end{equation} 
The numbers of independent components in 
$\ket{v_{S^z}}$ are given by the trinomial coefficients $T_{N,S^z}$, 
defined by 
\begin{equation}
(x^{-1}+1+x)^N=\sum_{k=-N}^{N} T_{N,k} x^{k}.
\end{equation} 
They provide the norm of the ground states 
\begin{equation}
\bracket{v_{S^z}}{v_{S^z}}=T_{N,S^z}.
\end{equation}
Details about trinomial coefficients can be found, e.g., in \cite{A027907}.

As a next step, we address structure of operators acting 
in the subspace of the ground state vectors, 
which can be regarded  
raising and lowering operators, in 
analogy with the Heisenberg XXX spin chain \cite{F-96}. We denote them by 
$\Sigma^{+}$ and $\Sigma^{-}$, respectively. 
More exactly, these operators should satisfy  
\begin{equation}\label{Sigmavec}
\Sigma^\pm\ket{v_{S^z}}
=
\begin{cases}
c_\pm(S^z)\ket{v_{S^z\pm1}}, & S^z\ne \pm N,
\\
0, & S^z=\pm N,
\end{cases}
\end{equation}
where $c_\pm(S^z)\ne 0$ are some constants.   
Furthermore, these operators should commute with the Hamiltonian,
\begin{equation}\label{SpmH}
\left[\Sigma^\pm,\mathrm{H}^\text{periodic}\right]=0.
\end{equation}
The following formulas turn out to be satisfying the above conditions.

\begin{conjecture}
There exist raising and lowering 
operators satisfying \eqref{Sigmavec} and \eqref{SpmH},  
and they are given by 
\begin{equation}\label{Sigmapmsum}
\Sigma^\pm=
\sum_{\substack{r_1,\dots,r_N\in\{-2,-1,0,1,2\}\\ r_1+\dots+r_N=\pm 1}} s_1^{r_1}\cdots s_N^{r_N}
\end{equation}
with the following notation:
\begin{equation}
s_i^{0}\equiv I, \qquad 
s_i^{\pm 1}\equiv s_i^{\pm}, \qquad 
s_i^{\pm 2}\equiv (s_i^{\pm})^2.
\end{equation}
Equivalently,  
\begin{equation}\label{Sigmapm}
\Sigma^\pm=\res_{\lambda=0}\,\prod_{i=1}^N 
\Bigl(\lambda^{-2} (s_i^{\pm})^2+\lambda^{-1} s_i^{\pm} 
+I +\lambda s_i^{\mp}+\lambda^2 (s_i^{\mp})^2\Bigr).
\end{equation}  
In these formulas, $I$ is the identity operator and  
$s_i^\pm$ are local spin-$1$ operators defined in \eqref{spin1rep} 
and \eqref{spmsz}. 
\end{conjecture}

Operators $\Sigma^\pm$ are essentially non-local and 
cannot be expressed in terms 
of other components of the total spin operator. 
When represented as matrices, their entries are just $0$ and $1$; they are not 
strictly upper or lower triangle matrices although nilpotent, 
$(\Sigma^\pm)^{2N+1}=0$.   
The number of terms in \eqref{Sigmapmsum} for $N=1,2,3,4,5,\ldots$ 
is given by the corresponding element in the 
sequence $1,4,18,80,365,\dots$, see \cite{A104631}.

The further query concerns which algebra the operators $\Sigma^\pm$ generate.
Apparently, one can expect appearance here a 
complex semi-simple Lie algebra. Such an algebra 
(see, e.g., \cite{K-02,IR-18}) is  
generated by $k$ triples, called  Chevalley generators, 
$e_i, f_i, h_i$, $i=1,\dots, k$, where $k$ is a rank,  
which satisfy so-called Serre relations  
\begin{equation}\label{Serre}
\begin{gathered}
[h_i,h_j]=0,
\qquad [e_i,f_i]=h_i,
\qquad
[e_i,f_j]=0,\quad i\ne j,
\\
[h_i,e_j]=A_{ij}e_j, \qquad [h_i,f_j]=-A_{ij}f_j,
\\ 
(\ad e_i)^{1-A_{ij}}e_j=(\ad f_i)^{1-A_{ij}} f_j=0,\quad i\ne j,
\end{gathered}
\end{equation}
where $A_{ij}$ are entries of the Cartan matrix and 
$\ad a \equiv [a,\cdot\,]$. 
The algebra is 
fixed up to an isomorphism by the Cartan matrix, which in our case 
need to be computed. 

To accomplish this task, one can introduce, in addition to 
$\Sigma^\pm$, operator 
$\Sigma^z=[\Sigma^{+},\Sigma^{-}]$ and further construct recursively triples 
$\Lambda^\pm=\pm\left[\Sigma^z,\Sigma^{\pm}\right]$,
$\Lambda^z=\left[\Lambda^{+},\Lambda^{-}\right]$, etc.  
We find that operators $\Sigma^z,\Lambda^z,\dots$ form an
abelian subalgebra, hence elements $h_i$ 
can be expected to be linear 
combinations of these operators. 
The rank of the algebra is identified by how many operators 
$\Sigma^z,\Lambda^z,\dots$ are linearly independent. 
We construct elements $e_i$ (simple roots)  
as linear combinations of operators $\Sigma^{+},\Lambda^{+},\dots$, and 
elements $f_i$ as 
linear combinations of operators $\Sigma^{-},\Lambda^{-},\dots$, with the same 
coefficients. Calculations towards fulfilling 
Serre relations \eqref{Serre}, with the additional requirement 
that elements $h_i$ are normalized canonically, $A_{ii}=2$, 
have led us to the following observation.  

\begin{conjecture}
The operators $\Sigma^\pm$ generate 
algebra \eqref{Serre} of rank $k=N$ with 
\begin{equation}
A=
\begin{pmatrix}
2 & -1 & 0 & 0 & 0 &\dots &0\\
-1 & 2 & -1 & 0 & 0 & \dots & 0\\
0 & -1 & 2 & -1 & 0 & \dots & 0\\
\hdotsfor{7}\\
0& \dots & 0 &-1 & 2 & -1 & 0\\
0& \dots & 0 & 0 & -1 & 2 & -1\\
0& \dots & 0 & 0 & 0& -2 & 2 \\
\end{pmatrix},
\end{equation}
i.e., the Lie algebra $C_N=\mathfrak{sp}_{2N}$.
\end{conjecture}

The last question we would like to address is a 
role of the operator $\mathrm{S}^z$ within the observed 
symmetry algebra. Noting that 
$[\mathrm{S}^z,\Sigma^\pm]=\pm \Sigma^\pm$, one can expect 
that $\mathrm{S}^z$ is expressed in terms of the Cartan subalgebra elements. 
It turns out that such a construction is indeed 
possible but involves one more symmetry generator. 

\begin{conjecture}
Operator $\mathrm{S}^z$ can be given as a sum
\begin{equation}\label{SzN}
\mathrm{S}^z=p+\sum_{i=1}^N\alpha_i h_i,
\end{equation}  
where $h_i$ are the Cartan subalgebra elements, $\alpha_i$ are 
some coefficients, and 
$p$ is an operator commuting with the operators $\Sigma^{\pm}$ and hence 
with all Chevalley generators, $[p,e_i]=[p,f_i]=[p,h_i]$.   
\end{conjecture}

In total, from  our considerations here one can conclude that 
an algebra $\mathcal{A}$ of symmetries of the periodic Motzkin chain 
Hamiltonian \eqref{Hpbc} is 
\begin{equation}
\mathcal{A}=\mathfrak{gl}_1\oplus \mathfrak{gl}_1\oplus \mathfrak{sp}_{2N},
\end{equation}
where one
$\mathfrak{gl}_1$ term is generated by the cyclic shift operator \eqref{Cop} 
and another one by the element $p$.  

\section{Examples}

Here we consider in detail examples of the periodic chain with 
two, three, and four sites. 

\subsection{Two-site chain}

The Hamiltonian of the two-site chain is  
\begin{equation}
\mathrm{H}^\mathrm{periodic}=\Pi+\mathrm{P} \Pi \mathrm{P},
\end{equation}
where $\mathrm{P}$ is the permutation operator. 
As a $9\times 9$ matrix, the Hamiltonian reads  
\begin{equation}
\mathrm{H}^\mathrm{periodic}=
\begin{pmatrix}
0 & 0 & 0 & 0 & 0 & 0 & 0 & 0 & 0 \\
 0 & 1 & 0 & -1 & 0 & 0 & 0 & 0 & 0 \\
 0 & 0 & \frac{1}{2} & 0 & -\frac{1}{2} & 0 & 0 & 0 & 0 \\
 0 & -1 & 0 & 1 & 0 & 0 & 0 & 0 & 0 \\
 0 & 0 & -\frac{1}{2} & 0 & 1 & 0 & -\frac{1}{2} & 0 & 0 \\
 0 & 0 & 0 & 0 & 0 & 1 & 0 & -1 & 0 \\
 0 & 0 & 0 & 0 & -\frac{1}{2} & 0 & \frac{1}{2} & 0 & 0 \\
 0 & 0 & 0 & 0 & 0 & -1 & 0 & 1 & 0 \\
 0 & 0 & 0 & 0 & 0 & 0 & 0 & 0 & 0 \\
\end{pmatrix}.
\end{equation}
The ground state ($H^\text{periodic}=0$) is five times 
degenerate. The independent states can be labeled by the eigenvalues
$S^z=0,\pm 1, \pm 2$ of the operator
\begin{equation}
\mathrm{S}^z=\diag(2,1,0,1,0,-1,0,-1,-2),
\end{equation}
and they read    
\begin{equation}
\ket{v_{2}}=
\begin{pmatrix}
1 \\ 0 \\ 0 \\ 0 \\ 0 \\ 0 \\ 0 \\ 0 \\ 0    
\end{pmatrix},\quad 
\ket{v_1}=
\begin{pmatrix}
0 \\ 1 \\ 0 \\ 1 \\ 0 \\ 0 \\ 0 \\ 0 \\ 0    
\end{pmatrix},\quad 
\ket{v_0}=
\begin{pmatrix}
0 \\ 0 \\ 1 \\ 0 \\ 1 \\ 0 \\ 1 \\ 0 \\ 0    
\end{pmatrix},\quad 
\ket{v_{-1}}=
\begin{pmatrix}
0 \\ 0 \\ 0 \\ 0 \\ 0 \\ 1 \\ 0 \\ 1 \\ 0    
\end{pmatrix},\quad 
\ket{v_{-2}}=
\begin{pmatrix}
0 \\ 0 \\ 0 \\ 0 \\ 0 \\ 0 \\ 0 \\ 0 \\ 1    
\end{pmatrix}, 
\end{equation}
or 
\begin{gather}
\ket{v_2}=\ket{\mathrm{uu}},\qquad
\ket{v_1}=\ket{\mathrm{uf}}+\ket{\mathrm{fu}},\qquad
\ket{v_0}=\ket{\mathrm{ud}}+\ket{\mathrm{ff}}+\ket{\mathrm{du}},
\\
\ket{v_{-1}}=\ket{\mathrm{fd}}+\ket{\mathrm{df}},\qquad
\ket{v_{-2}}=\ket{\mathrm{dd}}.
\end{gather}
Degeneracy of the ground state hints at existence of operators 
$\Sigma^\pm$ such that 
$\Sigma^\pm: \ket{v_{S^z}}\mapsto \ket{v_{S^z\pm1}}$  
and $\Sigma^\pm \ket{v_{\pm2}}=0$, and which, 
furthermore, commute with the Hamiltonian, 
$[\Sigma^\pm,\mathrm{H}^\text{periodic}]=0$.  
An easy guess leads us to the expression
\begin{equation}\label{Sigma-pm}
\Sigma^{+}=
\begin{pmatrix}
0 & 1 & 0 & 1 & 0 & 0 & 0 & 0 & 0 \\
 0 & 0 & 1 & 0 & 1 & 0 & 1 & 0 & 0 \\
 0 & 0 & 0 & 0 & 0 & 1 & 0 & 1 & 0 \\
 0 & 0 & 1 & 0 & 1 & 0 & 1 & 0 & 0 \\
 0 & 0 & 0 & 0 & 0 & 1 & 0 & 1 & 0 \\
 0 & 0 & 0 & 0 & 0 & 0 & 0 & 0 & 1 \\
 0 & 0 & 0 & 0 & 0 & 1 & 0 & 1 & 0 \\
 0 & 0 & 0 & 0 & 0 & 0 & 0 & 0 & 1 \\
 0 & 0 & 0 & 0 & 0 & 0 & 0 & 0 & 0 \\
\end{pmatrix},
\qquad
\Sigma^{-}=(\Sigma^{+})^\mathsf{T}.
\end{equation} 
In an operator form, 
\begin{equation}
\Sigma^\pm=s_1^\pm+s_2^\pm+(s_1^\pm)^2 s_2^\mp+s_1^\mp (s_2^\pm)^2, 
\end{equation}
where $s_j^\pm$ are local spin operators defined in \eqref{spin1rep} and 
\eqref{spmsz}. 

We further introduce operator
\begin{equation}\label{Sigmaz}
\Sigma^z=\left[\Sigma^{+},\Sigma^{-}\right],
\end{equation}
and from the explicit expression
\begin{equation}
\Sigma^{z}=
\begin{pmatrix}
2 & 0 & 0 & 0 & 0 & 0 & 0 & 0 & 0 \\
 0 & 2 & 0 & 2 & 0 & 0 & 0 & 0 & 0 \\
 0 & 0 & 0 & 0 & 0 & 0 & 0 & 0 & 0 \\
 0 & 2 & 0 & 2 & 0 & 0 & 0 & 0 & 0 \\
 0 & 0 & 0 & 0 & 0 & 0 & 0 & 0 & 0 \\
 0 & 0 & 0 & 0 & 0 & -2 & 0 & -2 & 0 \\
 0 & 0 & 0 & 0 & 0 & 0 & 0 & 0 & 0 \\
 0 & 0 & 0 & 0 & 0 & -2 & 0 & -2 & 0 \\
 0 & 0 & 0 & 0 & 0 & 0 & 0 & 0 & -2 \\
\end{pmatrix}
\end{equation}
we see that $\Sigma^z\ne \mathrm{S}^z$. Hence
$\Sigma^\pm$ and $\mathrm{S}^z$ do not span $\mathfrak{sl}_2$; a simple 
check also shows that $\Sigma^{\pm,z}$ do not span it neither. 

To study the algebra generated by operators \eqref{Sigma-pm}, we introduce 
operators $\Lambda^{\pm,z}$ by
\begin{equation}\label{Lambdapmz}
\Lambda^\pm=\pm\left[\Sigma^z,\Sigma^{\pm}\right],
\qquad
\Lambda^z=\left[\Lambda^{+},\Lambda^{-}\right].
\end{equation}
In particular, 
\begin{equation}
\Lambda^z=
\begin{pmatrix}
8 & 0 & 0 & 0 & 0 & 0 & 0 & 0 & 0 \\
 0 & 44 & 0 & 44 & 0 & 0 & 0 & 0 & 0 \\
 0 & 0 & 0 & 0 & 0 & 0 & 0 & 0 & 0 \\
 0 & 44 & 0 & 44 & 0 & 0 & 0 & 0 & 0 \\
 0 & 0 & 0 & 0 & 0 & 0 & 0 & 0 & 0 \\
 0 & 0 & 0 & 0 & 0 & -44 & 0 & -44 & 0 \\
 0 & 0 & 0 & 0 & 0 & 0 & 0 & 0 & 0 \\
 0 & 0 & 0 & 0 & 0 & -44 & 0 & -44 & 0 \\
 0 & 0 & 0 & 0 & 0 & 0 & 0 & 0 & -8 \\
\end{pmatrix}
\end{equation}
and so $[\Sigma^z,\Lambda^z]=0$, hence one 
can conclude that $\Sigma^z$ and $\Lambda^z$ are linear combinations
of the Cartan subalgebra elements. 
Indeed, introducing one more triple 
$\Phi^\pm=\pm[\Lambda^z,\Lambda^\pm]$, $\Phi^z=[\Phi^{+},\Phi^{-}]$, 
one can check that $\Phi^z$ is a linear combination 
of $\Sigma^z$ and $\Lambda^z$, hence the dimension of a 
Cartan subalgebra is $2$. To identify the algebra, we have to construct 
its simple roots and derive the Cartan matrix.
 
We will search elements $e_i$ as linear combinations of the 
operators $\Sigma^{+}$ and $\Lambda^{+}$: 
\begin{equation}
e_i=\rho_i\left(\Sigma^{+}+a_i\Lambda^{+}\right),\qquad i=1,2.
\end{equation}
Correspondingly, for elements $f_i$ we choose
\begin{equation}
f_i=\rho_i\left(\Sigma^{-}+a_i\Lambda^{-}\right),\qquad i=1,2.
\end{equation}
Commutation relations $[e_1,f_2]=0$ and $[e_2,f_1]=0$ yield two 
linear equations for $a_1$ and $a_2$, from which we find
\begin{equation}
a_1=-\frac{1}{4},\qquad a_2=\frac{1}{2}.
\end{equation}
The Cartan subalgebra elements can be constructed 
by setting $h_i=[e_i,f_i]$, $i=1,2$. Requiring 
$[h_i,e_i]=2e_i$, i.e.,  
the diagonal entries of the Cartan matrix are set to be 
$A_{11}=A_{22}=2$, we obtain 
\begin{equation}
\rho_1=\frac{\sqrt{2}}{3},\qquad 
\rho_2=\frac{1}{3 \sqrt{3}}.
\end{equation}
As a result, we find 
\begin{align}
e_1&=\frac{1}{\sqrt{2}}
\begin{pmatrix}
0 & 1 & 0 & 1 & 0 & 0 & 0 & 0 & 0 \\
 0 & 0 & 0 & 0 & 0 & 0 & 0 & 0 & 0 \\
 0 & 0 & 0 & 0 & 0 & 0 & 0 & 0 & 0 \\
 0 & 0 & 0 & 0 & 0 & 0 & 0 & 0 & 0 \\
 0 & 0 & 0 & 0 & 0 & 0 & 0 & 0 & 0 \\
 0 & 0 & 0 & 0 & 0 & 0 & 0 & 0 & 1 \\
 0 & 0 & 0 & 0 & 0 & 0 & 0 & 0 & 0 \\
 0 & 0 & 0 & 0 & 0 & 0 & 0 & 0 & 1 \\
 0 & 0 & 0 & 0 & 0 & 0 & 0 & 0 & 0 \\
\end{pmatrix},\qquad f_1=e_1^\mathsf{T},
\\
e_2&=\frac{1}{\sqrt{3}}
\begin{pmatrix}
0 & 0 & 0 & 0 & 0 & 0 & 0 & 0 & 0 \\
 0 & 0 & 1 & 0 & 1 & 0 & 1 & 0 & 0 \\
 0 & 0 & 0 & 0 & 0 & 1 & 0 & 1 & 0 \\
 0 & 0 & 1 & 0 & 1 & 0 & 1 & 0 & 0 \\
 0 & 0 & 0 & 0 & 0 & 1 & 0 & 1 & 0 \\
 0 & 0 & 0 & 0 & 0 & 0 & 0 & 0 & 0 \\
 0 & 0 & 0 & 0 & 0 & 1 & 0 & 1 & 0 \\
 0 & 0 & 0 & 0 & 0 & 0 & 0 & 0 & 0 \\
 0 & 0 & 0 & 0 & 0 & 0 & 0 & 0 & 0 \\
\end{pmatrix},\qquad f_2=e_2^\mathsf{T}.
\end{align}
We also have 
\begin{equation}
h_1=
\begin{pmatrix}
1 & 0 & 0 & 0 & 0 & 0 & 0 & 0 & 0 \\
0 & -\frac{1}{2} & 0 & -\frac{1}{2} & 0 & 0 & 0 & 0 & 0 \\
0 & 0 & 0 & 0 & 0 & 0 & 0 & 0 & 0 \\
0 & -\frac{1}{2} & 0 & -\frac{1}{2} & 0 & 0 & 0 & 0 & 0 \\
0 & 0 & 0 & 0 & 0 & 0 & 0 & 0 & 0 \\
0 & 0 & 0 & 0 & 0 & \frac{1}{2} & 0 & \frac{1}{2} & 0 \\
0 & 0 & 0 & 0 & 0 & 0 & 0 & 0 & 0 \\
0 & 0 & 0 & 0 & 0 & \frac{1}{2} & 0 & \frac{1}{2} & 0 \\
0 & 0 & 0 & 0 & 0 & 0 & 0 & 0 & -1 \\
\end{pmatrix}
\end{equation}
and
\begin{equation}
h_2=
\begin{pmatrix}
0 & 0 & 0 & 0 & 0 & 0 & 0 & 0 & 0 \\
0 & 1 & 0 & 1 & 0 & 0 & 0 & 0 & 0 \\
0 & 0 & 0 & 0 & 0 & 0 & 0 & 0 & 0 \\
0 & 1 & 0 & 1 & 0 & 0 & 0 & 0 & 0 \\
0 & 0 & 0 & 0 & 0 & 0 & 0 & 0 & 0 \\
0 & 0 & 0 & 0 & 0 & -1 & 0 & -1 & 0 \\
0 & 0 & 0 & 0 & 0 & 0 & 0 & 0 & 0 \\ 
0 & 0 & 0 & 0 & 0 & -1 & 0 & -1 & 0 \\ 
0 & 0 & 0 & 0 & 0 & 0 & 0 & 0 & 0 \\
\end{pmatrix}.
\end{equation}
From these expressions it follows that 
\begin{equation}\label{h1e2}
[h_1,e_2]=-e_2,\qquad [h_2,e_1]=-2e_1,
\end{equation}
and also
\begin{equation}\label{e1e1e2}
[e_1,[e_1,e_2]]=0,\qquad [e_2,[e_2,[e_2,e_1]]]=0.
\end{equation}
Similar relations hold for the elements $f_i$.  
Relations \eqref{h1e2} and \eqref{e1e1e2} 
imply that $A_{12}=-1$ and $A_{21}=-2$. Hence,
\begin{equation}
A=
\begin{pmatrix}
2 & -1 \\ -2 & 2  
\end{pmatrix}.
\end{equation}
This is the Cartan matrix of the Lie algebra $C_2$.
Note that if would exchange in the notation 
$e_1\leftrightarrow e_2$, $f_1\leftrightarrow f_2$, 
and $h_1\leftrightarrow h_2$, then 
we will obtain the transposed Cartan matrix $A^\mathsf{T}$, 
which corresponds to the algebra $B_2$. This is a well known equivalence 
$B_2=C_2$.  

Let us return to the operator $\mathrm{S}^z$. It can be easily seen that 
$[\mathrm{S}^z,\Sigma^\pm]=\pm \Sigma^\pm$ and one may wonder how 
$\mathrm{S}^z$ is expressed in terms of the Cartan subalgebra elements. 
Direct inspection shows that the operator 
\begin{equation}
p:=\mathrm{S}^z-\frac{7}{6}\Sigma^z+\frac{1}{24}\Lambda^z
\end{equation}
or, explicitly, 
\begin{equation}
p=\frac{1}{2}
\begin{pmatrix}
 0 & 0 & 0 & 0 & 0 & 0 & 0 & 0 & 0 \\
 0 & 1 & 0 & -1 & 0 & 0 & 0 & 0 & 0 \\
 0 & 0 & 0 & 0 & 0 & 0 & 0 & 0 & 0 \\
 0 & -1 & 0 & 1 & 0 & 0 & 0 & 0 & 0 \\
 0 & 0 & 0 & 0 & 0 & 0 & 0 & 0 & 0 \\
 0 & 0 & 0 & 0 & 0 & -1 & 0 & 1 & 0 \\
 0 & 0 & 0 & 0 & 0 & 0 & 0 & 0 & 0 \\
 0 & 0 & 0 & 0 & 0 & 1 & 0 & -1 & 0 \\
 0 & 0 & 0 & 0 & 0 & 0 & 0 & 0 & 0 \\
\end{pmatrix}
\end{equation}
is commuting with $\Sigma^\pm$ and hence it commutes with 
all elements of the algebra, i.e., $[p,e_i]=[p,f_i]=[p,h_i]=0$, $i=1,2$.   
The element $p$ is thus a central element in the symmetry algebra of the Hamiltonian. For the operator $\mathrm{S}^z$ 
the following representation is valid: 
\begin{equation}\label{Sz2}
\mathrm{S}^z=p+2h_1+\frac{3}{2}h_2. 
\end{equation}
We thus see, that it is given not only in terms 
of the Cartan subalgebra elements, 
but also contains the generating element of a one-dimensional center.

\subsection{Three-site chain}

The ground state vectors are given by (see 
\figurename~\ref{fig-ThreeSiteGS}):
\begin{gather}
\ket{v_{3}}=\ket{\mathrm{uuu}},\qquad 
\ket{v_{2}}=\ket{\mathrm{uuf}}+\ket{\mathrm{ufu}}+\ket{\mathrm{fuu}},
\\
\ket{v_{1}}=\ket{\mathrm{ffu}}+\ket{\mathrm{fuf}}+\ket{\mathrm{uff}}
+\ket{\mathrm{udu}}+\ket{\mathrm{uud}}+\ket{\mathrm{duu}},
\\
\ket{v_{0}}=\ket{\mathrm{fff}}+\ket{\mathrm{udf}}+\ket{\mathrm{fud}}
+\ket{\mathrm{ufd}}+\ket{\mathrm{duf}}+\ket{\mathrm{fdu}}
+\ket{\mathrm{dfu}},
\\
\ket{v_{-1}}=\ket{\mathrm{ffd}}+\ket{\mathrm{fdf}}+\ket{\mathrm{dff}}
+\ket{\mathrm{dud}}+\ket{\mathrm{duu}}+\ket{\mathrm{udd}},
\\
\ket{v_{-2}}=\ket{\mathrm{ddf}}+\ket{\mathrm{dfd}}+\ket{\mathrm{fdd}},
\qquad
\ket{v_{-3}}=\ket{\mathrm{ddd}}.
\end{gather}

The operators $\Sigma^\pm$ in this case have the following form:
\begin{multline}
\Sigma^\pm=s_1^\pm+s_2^\pm+s_3^\pm
+(s_1^\pm)^2 s_2^\mp+s_1^\mp (s_2^\pm)^2
+(s_1^\pm)^2 s_3^\mp+s_1^\mp (s_3^\pm)^2
+(s_2^\pm)^2 s_3^\mp+s_2^\mp (s_3^\pm)^2
\\
+s_1^\pm s_2^\pm s_3^\mp+s_1^\pm s_2^\mp s_3^\pm+s_1^\mp s_2^\pm s_3^\pm
+s_1^\pm (s_2^\pm)^2 (s_3^\mp)^2+(s_1^\pm)^2 s_2^\pm (s_3^\mp)^2
\\
+(s_1^\pm)^2 (s_2^\mp)^2 s_3^\pm+s_1^\pm (s_2^\mp)^2 (s_3^\pm)^2
+(s_1^\mp)^2 (s_2^\pm)^2 s_3^\pm+(s_1^\mp)^2 s_2^\pm (s_3^\pm)^2. 
\end{multline}
To study the algebra generated by $\Sigma^\pm$,
we introduce operator $\Sigma^z$ by \eqref{Sigmaz}, operators 
$\Lambda^{\pm,z}$ by \eqref{Lambdapmz}, and one more 
triple of operators $\Phi^{\pm,z}$ by
\begin{equation}\label{Phipmz}
\Phi^\pm=\pm \left[\Lambda^z,\Lambda^{\pm}\right],\qquad
\Phi^z=\left[\Phi^{+},\Phi^{-}\right].
\end{equation}
The elements $e_i$ can be searched in the form  
\begin{equation}
e_i=\rho_i\left(\Sigma^{+}+a_i\Lambda^{+}+b_i\Phi^{+}\right),
\qquad i=1,2,3,
\end{equation}
and, correspondingly, elements $f_i$ in the form 
\begin{equation}
f_i=\rho_i\left(\Sigma^{-}+a_i\Lambda^{-}+b_i\Phi^{-}\right),
\qquad i=1,2,3.
\end{equation}
Commutation relations $[e_i,f_j]=0$, $i\ne j$, fix $a_i$ and $b_i$ to be 
given as  
\begin{gather}
a_1=\frac{1081}{29\,628},\qquad 
a_2=\frac{277}{3456},\qquad 
a_3=\frac{581}{7038}, 
\\
b_1=-\frac{11}{3\,199\,824},\qquad
b_2=-\frac{1}{186\,624},\qquad
b_3=-\frac{1}{760\,104}.
\end{gather}
For the Cartan subalgebra elements we set  
$h_i=[e_i,f_i]$, $i=1,2,3$. Requiring 
$[h_i,e_i]=2e_i$, i.e., that  
the diagonal entries of the Cartan matrix are  
$A_{11}=A_{22}=A_{33}=2$, we obtain 
\begin{equation}
\rho_1=\frac{1646}{885\sqrt{3}},\qquad 
\rho_2=\frac{64\sqrt{2}}{295},\qquad
\rho_3=\frac{391}{885 \sqrt{21}}.
\end{equation}
Thus constructed simple roots satisfy 
\begin{gather}
[h_1,e_2]=-e_2,\qquad 
[h_1,e_3]=0,\qquad
[h_2,e_1]=-e_1,\qquad
[h_2,e_3]=-e_3,
\\ 
[h_3,e_1]=0,\qquad [h_3,e_2]=-2e_2.
\end{gather}
One can also verify that
\begin{gather}
[e_1,[e_1,e_2]]=0,\qquad [e_1,e_3]=0,
\qquad [e_2,[e_2,e_1]]=0,
\\
[e_2,[e_2,e_3]]=0,
\qquad [e_3,[e_3,[e_3,e_2]]]=0.
\end{gather}
Hence, 
\begin{equation}
A=
\begin{pmatrix}
2 & -1 & 0 \\
-1 & 2 & -1 \\
0 & -2 & 2  
\end{pmatrix}.
\end{equation}

A search for central element $p$ constructed as a linear combination 
of the operators $\mathrm{S}^z$, $\Sigma^z$, $\Lambda^z$, and $\Phi^z$
by solving, e.g., the relation $[p,\Sigma^{+}]=0$ yields
\begin{equation}
p=\mathrm{S}^z-\frac{792\,749}{3\,106\,467} \Sigma^z 
-\frac{1\,302\,389}{251\,623\,827} \Lambda^z 
+\frac{61}{586\,880\,636\,256} \Phi^z.
\end{equation}
This result, equivalently obtained in terms of elements of the Cartan 
subalgebra, implies  
\begin{equation}\label{Sz3}
\mathrm{S}^z=p+3h_1+5h_2+3h_3.
\end{equation}

\subsection{Four-site chain}  
In this case we give details only for construction 
of the Cartan matrix and present the result for the operator $\mathrm{S}^z$.  

We start with the operators $\Sigma^\pm$, for which 
the general formula \eqref{Sigmapm} works as expected, i.e., 
a check shows that \eqref{Sigmavec} and \eqref{SpmH} hold.
We define $\Sigma^z$ by \eqref{Sigmaz}, $\Lambda^{\pm,z}$ by \eqref{Lambdapmz},
$\Phi^{\pm,z}$ by \eqref{Phipmz}, and also introduce operators
\begin{equation}
\Omega^\pm=\pm \left[\Phi^z,\Phi^{\pm}\right],\qquad
\Omega^z=\left[\Omega^{+},\Omega^{-}\right].
\end{equation}
We search simple roots in the form
\begin{equation}
e_i=\rho_i\left(\Sigma^{+}+a_i\Lambda^{+}+b_i\Phi^{+}+c_i\Omega^{+}\right),
\qquad i=1,\ldots,4
\end{equation}
and, similarly, 
\begin{equation}
f_i=\rho_i\left(\Sigma^{-}+a_i\Lambda^{-}+b_i\Phi^{-}+c_i\Omega^{-}\right),
\qquad i=1,\ldots,4.
\end{equation}
Commutation relations $[e_i,f_j]=0$, $i\ne j$, are fulfilled with 
\begin{align}
a_1&= \frac{105\,625\,140\,496\,014\,730\,841\,477}
{7\,703\,529\,626\,668\,586\,930\,816\,688},
\\
a_2&=\frac{415\,175\,982\,533\,783\,376\,793}
{13\,752\,186\,821\,722\,991\,129\,796},
\\
a_3&=\frac{32\,936\,728\,012\,334\,124\,913\,399}
{1\,363\,174\,534\,869\,932\,976\,556\,176},
\\
a_4&=\frac{21\,741\,465\,949\,931\,994\,477\,137}
{904\,173\,010\,239\,198\,188\,108\,928},
\\
b_1&=-\frac{5\,256\,682\,134\,946\,428\,299}
{1\,365\,302\,481\,526\,494\,176\,046\,280\,704},
\\
b_2&=-\frac{3\,923\,011\,779\,201\,308\,513}
{1\,013\,921\,229\,991\,992\,690\,017\,599\,488},
\\
b_3&=-\frac{9\,024\,272\,054\,124\,165\,191}
{348\,972\,680\,926\,702\,841\,998\,381\,056},
\\
b_4&=-\frac{18\,259\,103\,029\,394\,551\,109}
{694\,404\,871\,863\,704\,208\,467\,656\,704},
\\
c_1&=\frac{326\,351}
{148\,888\,835\,146\,389\,016\,342\,758\,102\,889\,660\,416},
\\
c_2&=-\frac{74\,917}
{8\,505\,387\,741\,280\,669\,815\,423\,155\,205\,832\,704},
\\
c_3&=\frac{5311}{60\,987\,396\,312\,566\,393\,208\,549\,069\,029\,376},
\\
c_4&=-\frac{581\,743}
{5\,825\,090\,263\,354\,844\,032\,785\,452\,768\,428\,032}.
\end{align}
Relations $[h_i,e_i]=2e_i$, $i=1,\ldots,4$, yield
\begin{align}
\rho_1&=\frac{206\,344\,543\,571\,480\,007\,075\,447}
{217\,682\,719\,003\,513\,150\,677\,430},
\\
\rho_2&=\frac{3\,929\,196\,234\,777\,997\,465\,656}
{21\,768\,271\,900\,351\,315\,067\,743}\sqrt{\frac{2}{5}},
\\
\rho_3&=\frac{4\,057\,067\,068\,065\,276\,715\,941}
{108\,841\,359\,501\,756\,575\,338\,715 \sqrt{10}},
\\
\rho_4&=\frac{672\,747\,775\,475\,593\,889\,962}
{108\,841\,359\,501\,756\,575\,338\,715}\sqrt{\frac{2}{19}}.
\end{align} 
Thus constructed simple roots satisfy
\begin{gather}
[h_1,e_2]=-e_2,\qquad 
[h_1,e_3]=[h_1,e_4]=0,
\\
[h_2,e_1]=-e_1,\qquad
[h_2,e_3]=-e_3,\qquad
[h_2,e_4]=0,
\\ 
[h_3,e_1]=0,\qquad 
[h_3,e_2]=-e_2,\qquad
[h_3,e_4]=-e_4,
\\
[h_4,e_1]=[h_4,e_2]=0,\qquad
[h_4,e_3]=-2e_3.
\end{gather}
One can also check that 
\begin{gather}
[e_1,[e_1,e_2]]=0,\qquad [e_1,e_3]=0, \qquad [e_1,e_4]=0,
\\
[e_2,[e_2,e_1]]=0,\qquad [e_2,[e_2,e_3]]=0, \qquad [e_2,e_4]=0,
\\
[e_3,[e_3,e_2]]=0,\qquad 
[e_3,[e_3,e_4]]=0,
\\
[e_4,[e_4,[e_4,e_3]]]=0.
\end{gather}
Hence, 
\begin{equation}
A=
\begin{pmatrix}
2 & -1 & 0 & 0 \\
-1 & 2 & -1 & 0 \\
0 & -1 & 2 & -1 \\
0 & 0 & -2 & 2  
\end{pmatrix}.
\end{equation}

A search for central element $p$ constructed as a linear combination 
of the operators $\mathrm{S}^z$, $\Sigma^z$, $\Lambda^z$, $\Phi^z$,
and $\Omega^z$ by solving, e.g., the relation $[p,\Sigma^{+}]=0$ 
leads to a formula with enormous coefficients. 
Using the Cartan subalgebra elements we  
find that the result appears pretty simple:
\begin{equation}\label{Sz4}
\mathrm{S}^z=p+ 4h_1 + 7h_2 + 9h_3 + 5h_4.
\end{equation}

\section{Conclusion}

In this paper, we have considered the Motzkin spin chain 
with periodic boundary conditions and provided several conjectures concerning 
the structure of the ground state space and symmetries of the Hamiltonian. 
We find that the symmetry algebra of the model is quite rich, that can be 
interpreted as a signal of quantum integrability. Nevertheless, 
to claim that the Motzkin chain is a quantum integrable model 
it should be identified as such among solutions of the Yang--Baxter equation.
This task can be regarded as of primary interest, since it
would provide extremely powerful mathematical tools for 
computing various quantities of interest, such as correlation functions.

Here we briefly mention more specific directions for further research, 
in particular those aimed at proving our findings.
Conjecture~1 states that the ground state of the periodic chain with $N$ sites
is $(2N+1)$-fold degenerate, 
with independent states distinguished by the eigenvalue of 
the third component of total spin 
operator, and, moreover, that these states admit an
interpretation in terms of paths, similar to the Motzkin paths. 
A possible way to prove it could consist in generalizing  
the method developed in the context of the Fredkin spin chain \cite{SK-17},  
where it was shown that the degenerate ground states 
in the case of periodic boundary conditions can be 
described in terms of suitably defined classes of Dyck paths.

Concerning Conjecture~2, which provides an explicit expression 
for raising and lowering operators commuting with the Hamiltonian, 
we see no simple recipe for proving it except establishing underlying 
quantum integrable structures. However, a more mathematically 
sensible problem can be easily formulated: Is it possible to prove 
Conjecture~3 starting from 
the formulas provided in Conjecture~2 
for these operators? Definitely, this is a  
representation theory problem, and the question is about 
which representation (obviously, 
a reducible one) of $\mathfrak{sp}_{2N}$ is involved here. 
An answer to this question may also lead to a proof of Conjecture~4, 
about the central extension.

Another intriguing problem is related to combinatorics. 
Inspecting \eqref{Sz2}, \eqref{Sz3}, and \eqref{Sz4} one may be tempted to 
conjecture that the coefficients $\alpha_i$ in \eqref{SzN} 
are positive integers, except 
the two-site case which is in fact somewhat special. 
If they indeed appear to be positive integers, then one may next
wonder whether they admit any combinatorial interpretation, just like 
other objects we meet in our study here, such as 
the numbers of the ground states components given by 
the trinomial coefficients, 
and the numbers of terms in the operators $\Sigma^\pm$ given by the 
sequence $1,4,18,80,365,\dots$.

\section*{Acknowledgments}

The author is indebted for useful discussions to  
Nikolai M. Bogoliubov and Vitaly O. Tarasov.
This work is supported by Russian Science Foundation 
grant \#23-11-00311.

\bibliography{sympmot_bib}

\begin{bibdiv}
\begin{biblist}

\bib{BCMNS-12}{article}{
      author={Bravyi, S.},
      author={Caha, L.},
      author={Movassagh, R.},
      author={Nagaj, D.},
      author={Shor, P.},
       title={Criticality without frustration for quantum spin-1 chains},
        date={2012},
     journal={Phys. Rev. Lett.},
      volume={109},
       pages={207202 (5 pp.)},
      eprint={1203.5801},
         doi={10.1103/PhysRevLett.109.207202},
}

\bib{MS-16}{article}{
      author={Movassagh, R.},
      author={Shor, P.~W.},
       title={Supercritical entanglement in local systems: {C}ounterexample to
  the area law for quantum matter},
        date={2016},
     journal={Proc. Natl. Acad. Sci.},
      volume={113},
       pages={13278–13282},
      eprint={1408.1657},
         doi={10.1073/pnas.1605716113},
}

\bib{SK-17}{article}{
      author={Salberger, O.},
      author={Korepin, V.},
       title={Entangled spin chain},
        date={2017},
     journal={Rev. Math. Phys.},
      volume={29},
       pages={1750031 (20 pp.)},
      eprint={1605.03842},
         doi={10.1142/S0129055X17500313},
}

\bib{AAZK-19}{article}{
      author={Alexander, R.~N.},
      author={Ahmadain, A.},
      author={Zhang, Zh.},
      author={Klich, I.},
       title={Exact rainbow tensor networks for the colorful {M}otzkin and
  {F}redkin spin chains},
        date={2019},
     journal={Phys. Rev. B},
      volume={100},
       pages={214430 (6 pp.)},
      eprint={1811.11974},
         doi={10.1103/PhysRevB.100.214430},
}

\bib{CBG-24}{article}{
      author={Causer, L.},
      author={Banuls, M.~C.},
      author={Garrahan, J.~P.},
       title={Nonthermal eigenstates and slow relaxation in quantum {F}redkin
  spin chains},
        date={2024},
     journal={Phys. Rev. B},
      volume={110},
       pages={134322 (15 pp.)},
      eprint={2403.03986},
         doi={10.1103/PhysRevB.110.134322},
}

\bib{MGM-24}{article}{
      author={Menon, V.},
      author={Gu, A.},
      author={Movassagh, R.},
       title={Symmetries, correlation functions, and entanglement of general
  quantum {M}otzkin spin-chains},
      eprint={2408.16070},
}

\bib{TTF-83}{article}{
      author={Tarasov, V.~O.},
      author={Takhtadzhyan, L.~A.},
      author={Faddeev, L.~D.},
       title={Local {H}amiltonians for integrable quantum models on a lattice},
        date={1983},
     journal={Theor. Math. Phys.},
      volume={57},
       pages={1059\ndash 1073},
         doi={10.1007/BF01018648},
}

\bib{S-82}{article}{
      author={Sklyanin, E.~K.},
       title={Some algebraic structures connected with the {Y}ang--{B}axter
  equation},
        date={1982},
     journal={Funct. Anal. Its Appl.},
      volume={16},
       pages={263\ndash 270},
         doi={10.1007/BF01077848},
}

\bib{KBI-93}{book}{
      author={Korepin, V.~E.},
      author={Bogoliubov, N.~M.},
      author={Izergin, A.~G.},
       title={{Quantum Inverse Scattering Method and Correlation Functions}},
   publisher={Cambridge University Press},
     address={Cambridge},
        date={1993},
}

\bib{S-20}{article}{
      author={Slavnov, N.~A.},
       title={{Introduction to the nested algebraic Bethe ansatz}},
        date={2020},
     journal={SciPost Phys. Lect. Notes},
      volume={19},
      eprint={1911.12811},
         doi={10.21468/SciPostPhysLectNotes.19},
}

\bib{HSK-23}{article}{
      author={Hao, K.},
      author={Salberger, O.},
      author={Korepin, V.},
       title={Exact solution of the quantum integrable model associated with
  the {M}otzkin spin chain},
        date={2023},
     journal={J. High Energ. Phys.},
      volume={2023},
       pages={08009 (23 pp.)},
      eprint={2202.07647},
         doi={10.1007/JHEP08(2023)009},
}

\bib{K-23}{article}{
      author={Khachatryan, S.~A.},
       title={New series of multi-parametric solutions to {GYBE: Q}uantum gates
  and integrability},
        date={2023},
     journal={Nucl. Phys. B},
      volume={996},
       pages={116375},
      eprint={2308.03564},
         doi={10.1016/j.nuclphysb.2023.116375},
}

\bib{PK-24}{article}{
      author={Padmanabhan, P.},
      author={Korepin, V.},
       title={Solving the {Y}ang--{B}axter, tetrahedron and higher simplex
  equations using {C}lifford algebras},
        date={2024},
     journal={Nucl. Phys. B},
      volume={1007},
       pages={116664},
      eprint={2404.11501},
         doi={10.1016/j.nuclphysb.2024.116664},
}

\bib{SSPK-24}{article}{
      author={Singh, V.~K.},
      author={Sinha, A.},
      author={Padmanabhan, P.},
      author={Korepin, V.},
       title={Unitary tetrahedron quantum gates},
      eprint={2407.10731},
}

\bib{GBFG-24}{article}{
      author={Garkun, A.~S.},
      author={Barik, S.~K.},
      author={Fedorov, A.~K.},
      author={Gritsev, V.},
       title={New spectral-parameter dependent solutions of the
  {Y}ang--{B}axter equation},
      eprint={2401.12710},
}

\bib{A001006}{misc}{
      author={Sloane, N. J.~A.},
       title={Sequence {A001006}},
     address={The On-line Encyclopedia of Integer Sequences},
         url={https://oeis.org/A001006},
}

\bib{A027907}{misc}{
      author={Sloane, N. J.~A.},
       title={Sequence {A027907}},
     address={The On-line Encyclopedia of Integer Sequences},
         url={https://oeis.org/A027907},
}

\bib{F-96}{inproceedings}{
      author={Faddeev, L.~D.},
       title={How algebraic {B}ethe ansatz works for integrable model},
        date={1996},
   booktitle={{Les Houches School of Physics: Astrophysical Sources of
  Gravitational Radiation}},
       pages={149\ndash 219},
      eprint={hep-th/9605187},
}

\bib{A104631}{misc}{
      author={Sloane, N. J.~A.},
       title={Sequence {A104631}},
     address={The On-line Encyclopedia of Integer Sequences},
         url={https://oeis.org/A104631},
}

\bib{K-02}{book}{
      author={Knapp, A.~W.},
       title={{Lie Groups Beyond an Introduction}},
     edition={2},
      series={Progress in Mathematics},
   publisher={Birkh\"auser},
     address={Boston},
        date={2002},
      volume={140},
}

\bib{IR-18}{book}{
      author={Isaev, A.~P.},
      author={Rubakov, V.~A.},
       title={{Theory of Groups and Symmetries}},
    subtitle={{Finite groups, Lie Groups, and Lie Algebras}},
   publisher={World Scientific},
     address={Singapore},
        date={2018},
}

\end{biblist}
\end{bibdiv}
\end{document}